\documentclass[12pt]{article}
\usepackage{amsmath} 
\usepackage{graphicx}
\usepackage{rotating} 
\newcommand{\bmx}{\mbox{\boldmath $x$}}   
\newcommand{\bmg}{\mbox{\boldmath $g$}}
\newcommand{\bmc}{\mbox{\boldmath $c$}}

\newcommand{\bmzero}{\mbox{\boldmath $0$}}
\newcommand{\bmX}{\mbox{\boldmath $X$}}
\newcommand{\bmN}{\mbox{\boldmath $N$}}

\newcommand{\bmD}{\mbox{\boldmath $D$}}
\newcommand{\bg}{\mbox{\boldmath $\gamma$}}

\newcommand{\bmd}{\mbox{\boldmath $d$}}
\newcommand{\bb}{\mbox{\boldmath $\beta$}}

\newcommand{\bn}{\mbox{\boldmath $\eta$}}
\newcommand{\bt}{\mbox{\boldmath $\theta$}}

\newcommand{\bmb}{\mbox{\boldmath $b$}}

\newcommand{\bmZ}{\mbox{\boldmath $Z$}}
\newcommand{\bmz}{\mbox{\boldmath $z$}}
\newcommand{\bms}{\mbox{\boldmath $s$}}
\newcommand{\bmr}{\mbox{\boldmath $r$}}

\DeclareMathOperator*{\argmin}{arg\,min}

\begin{document}

\pagenumbering{gobble}

\begin{center}
\noindent  {\bf A Modified Net Reclassification Improvement Statistic}

\vskip 0.35in

\noindent {Glenn Heller}

Department of Epidemiology and Biostatistics,  

Memorial Sloan Kettering,  New York, NY 10017, U.S.A. \\ 
{\em $^*$email: hellerg@mskcc.org}

\end{center}

\vskip 0.25in

\noindent {\bf ABSTRACT} 
\newline The continuous net reclassification improvement (NRI) statistic is a popular model change measure that was developed to assess the incremental value of new factors in a risk prediction model. Two prominent statistical issues identified in the literature call the utility of this measure into question: (1) it is not a proper scoring function and (2) it has a high false positive rate when testing whether new factors contribute to the risk model. For binary response regression models, these subjects are interrogated and a modification of the continuous NRI, guided by the likelihood-based score residual, is proposed to address these issues.  Within a nested model framework, the modified NRI may be viewed as a distance measure between two risk models. An application of the modified NRI is illustrated using prostate cancer data.

\vskip 0.25in

\noindent {\bf KEYWORDS} 
\newline Binary response model, $L_1$ distance, Nested models, Proper score, Valid test

\newpage

\pagenumbering{arabic}
\setcounter{page}{1}

\noindent {\bf 1. INTRODUCTION} 

\noindent In the clinical setting, individual risk assessment is often derived through a regression model, which incorporates a combination of risk factors due to biological complexity. These risk models are used in forecasting future health outcomes of an individual such as treatment response or survival. The quality of the risk model, evaluated using statistical measures such as calibration, discrimination, explained variation, and likelihood based, reflects the level of confidence in the forecast (Gerds and Kattan 2021).  When the objective is to incorporate a new set of factors to an existing risk model, assessing the impact of these new factors on the forecast is critical. For binary response regression, a discrimination measure, the net reclassification improvement (NRI), is one statistic used for this evaluative process. The NRI, also referred to as the net reclassification index, was developed to ascertain whether the introduction of new risk factors move a model derived forecast in a direction consonant with the binary response outcome (Pencina et al. 2008). 

The NRI statistic has been criticized on numerous grounds. Two prevailing points of contention are: (1) it is not a proper scoring function and (2) it has a high false positive rate when testing whether the new factors contribute to the risk model, even in situations that include independent training and test datasets (Kerr et al. 2014 and Pepe et al. 2014, 2015). Despite of these critiques, the NRI is a popular statistic, and in the three-year time period 2019-2021, it was cited in PubMed over 800 times. The purpose of this work is to elucidate the methodology underlying these two concerns and to propose a likelihood guided modification to the NRI to rectify these issues.   

The NRI is defined through a series of nested regression models, where it is assumed that the existing factors alone $(\bmx)$, which includes a constant for the intercept term, or combined with new factors $(\bmz)$ are modeled as   
\begin{align} 
\begin{split}
\mbox{Pr}(Y=1) & = G(\beta^{\bullet}) \\
\mbox{Pr}(Y=1 |\bmx) & = G({\bb^0}^T \bmx) \\
\mbox{Pr}(Y=1 |\bmx, \bmz) & = G(\bb_0^T \bmx + \bg_0^T \bmz) 
\end{split}
\end{align}
where $Y$ is a binary outcome denoted as event ($Y=1$) or non-event ($Y=0$), $G$ is a monotone function representing the probability of an event, the base model risk score is ${\bb^0}^T \bmx$, the expanded model risk score is $\bb_0^T \bmx + \bg_0^T \bmz$, and for the constant model, $\pi_0 = G(\beta^{\bullet})$.  Throughout this work, random variables are represented with upper case, their observed copies are written in lower case, and vectors are indicated in bold.

The log-likelihood used to estimate the model parameters is 
\[ L(\bb, \bg) = \sum_i \left[y_i \log G(\bb^T \bmx_i + \bg^T \bmz_i) + (1-y_i) \log (1-G(\bb^T \bmx_i + \bg^T \bmz_i)) \right] ,\]
where $\{(y_i,\bmx_i, \bmz_i)\}, i=1, \ldots, n$ are independent identically distributed copies of $(Y, \bmX, \bmZ)$.  The maximum likelihood estimates from the three models are represented as: $\hat{\bt} = (\hat{\bb}, \hat{\bg})$, $\hat{\bt}^0 = (\hat{\bb}^0, \bmzero)$, and $\hat{\pi} = \bar{y}$, the observed proportion of events.

Historically, the NRI was developed under the assumption that the base model risk score could be placed in risk classification categories. It was a measure of whether the expanded model risk score, due to the addition of new factors, would move into higher risk categories for subjects with an event and into lower risk categories for subjects without an event. This framework, however, requires apriori clinically meaningful risk categories, which are often not apparent at the time of analysis, particularly in the early stage of model development. As a result, the continuous NRI was developed (Pencina et al. 2011) and it is this measure that is the focus of this work.

The population NRI is defined as
\begin{eqnarray*}
\lefteqn{\rho(\bt_0; \bt^0;\pi_0) = } \\
 & & 2 \left\{ \mbox{Pr}(\bb_0^T \bmX + \bg_0^T \bmZ \ge {\bb^0}^T \bmX |Y=1)  -  \mbox{Pr}(\bb_0^T \bmX + \bg_0^T \bmZ \ge {\bb^0}^T \bmX |Y=0) \right\} 
\end{eqnarray*} 
where $\bt_0 = (\bb_0, \bg_0)$ and $\bt^0 = (\bb^0,\bmzero)$. When multiplied by $1/2$, the population NRI is estimated as 
\begin{equation}  R_n(\hat{\bt}; \hat{\bt}^0;\hat{\pi}) = \left[n \bar{y} (1-\bar{y})\right]^{-1} \sum_i \left[y_i - \bar{y} \right] \left[I(\hat{\bb}^T \bmx_i +\hat{\bg}^T \bmz_i - \hat{\bb^0}^T \bmx_i > 0 ) - \frac{1}{2} \right] .\end{equation} 
Assuming at least one component of $\bmx$ is continuous, it can be asserted without loss of generality, that the indicator function can be extended as  
\begin{align}
\begin{split}
I(u > 0) = 1 & \ \ \ \mbox{if}  \ \ u > 0 \\
I(u > 0) = \frac{1}{2} & \ \ \ \mbox{if}  \ \ u = 0 \\
I(u > 0) = 0 & \ \ \ \mbox{if}  \ \ u < 0 .
\end{split}
\end{align}

Although the net reclassification improvement statistic is a frequently applied model change measure, its lack of propriety and high false positive rate are problematic. In Section 2, a modified NRI (mNRI) is developed that satisfies the concept of a proper change score, which adapts the proper scoring principle to model change measures (Pepe et al. 2015). Section 3 demonstrates that a smooth version of the mNRI provides a valid test procedure when the population NRI is zero. This result is established in the single sample and the independent training and test data case. In Section 4, a prostate cancer data example is used to illustrate these concepts and Section 5 contains a discussion.

\vskip 0.50in

\noindent {\bf 2. THE mNRI IS A PROPER CHANGE SCORE} 

\noindent For a correctly specified parametric risk model, a performance measure is a proper score if its expected value is minimized/maximized at the true model parameter value (Gneiting and Raftery, 2007). For example, the expected value of the Brier score applied to the expanded model 
\[ E[Y - G(\bb^T \bmX + \bg^T \bmZ)]^2 , \]
 is minimized at $(\bb, \bg) = (\bb_0, \bg_0)$.  
If a performance measure is not a proper score, then the analyst may find inconsistent parameter estimates that make the measure look better. Population performance measures such as the expected value of the area under the curve (AUC), the Brier score (BS), and Kullback-Leibler divergence (KL), are maximized/minimized at their true parameter values and therefore are proper scores. 
 
Proper scoring is more difficult to achieve for model change measures. Consider the case where a performance measure $M$ is applied separately to the expanded model and the base model, and the change measure is  
\[ \Delta M(\bmb, \bmg;\bmb^0) = M(\bmb, \bmg) - M(\bmb^0). \]
If the performance measure (M) is convex,   
\[ (\bb_0, \bg_0) = \argmin _{(\bmb, \bmg)} \ \ E[M(\bmb, \bmg)] \] 
\[ {\bb^0} = \argmin _{\bmb^0} \ \ E[M(\bmb^0)] , \] 
but the difference of two convex functions is not necessarily convex, and in general,
\[ (\bb_0, \bg_0,{\bb^0}) \ne \argmin_{(\bmb,\bmg,\bmb^0)} \ \ E[\Delta M(\bmb, \bmg;\bmb^0)] . \]

To adapt proper scoring to change measures, Pepe et al. (2015) orient the model parameter space so that the base model is evaluated at the true parameter ${\bb^0}$. In this setting, $\Delta M$ is termed a proper change score, since 
\[ (\bb_0, \bg_0) = \argmin _{(\bmb, \bmg)} \ \ E\left[\Delta M(\bmb, \bmg;{\bb^0}) \right] ,\] 
recreating the single model evaluation. The term proper change score is used here to acknowledge the adaptation of the proper scoring principle to change measures. Under this definition, $\Delta$AUC, $\Delta$BS, and $\Delta$KL are proper change scores. 

The NRI differs from other change measures because it is a statistic based on within subject change and not between model change as above. In addition, the statistic is composed of parameter estimates from three nested models.  As a result, it is not covered under the previous argument.  To satisfy the proper change score criterion, the NRI is modified
\[  T_n(\hat{\bt}; \hat{\bt}^0; \hat{\pi}) = \left[n \bar{y} (1-\bar{y})\right]^{-1} \sum_i r(\hat{\bb^0}^T\bmx_i) \left[I(\hat{\bb}^T \bmx_i +\hat{\bg}^T \bmz_i - \hat{\bb^0}^T \bmx_i > 0 ) - \frac{1}{2} \right] ,\]
which is constructed by replacing the constant model score residual $y - \bar{y}$ in (2) with the base model score residual $r(\hat{\bb^0}^T \bmx)$, where
\begin{equation} r({\bb^0}^T \bmx) = \left[\frac{\partial G({\bb^0}^T \bmx)}{\partial ({\bb^0}^T \bmx)}\right] \left[G({\bb^0}^T \bmx)(1-G({\bb^0}^T \bmx))\right]^{-1} \left[y_i - G({\bb^0}^T \bmx) \right] . 
\end{equation}
The modified NRI (mNRI) is closely akin to the maximum score statistic and the least absolute deviation statistic (Manski 1985, Horowitz 1998), which provide the framework for the derivation in Theorem 1. 

\vskip 0.30in   
 
\noindent {\bf Theorem 1. }  

\noindent Consider the mNRI scoring function derived from a single random variable, with the base and constant model parameters given
\[ T_1 (\bt; \bt^0; \pi_0) = [\pi_0 (1-\pi_0)]^{-1} r({\bb^0}^T \bmX) \left[I(\bb^T \bmX +\bg^T \bmZ - {\bb^0}^T \bmX > 0) - \frac{1}{2} \right] .\] 

The mNRI scoring function is a proper change score,
\[ E[ T_1(\bt_0;\bt^0;\pi_0)] \ge E[T_1(\bt;\bt^0;\pi_0)] \ \ \mbox{ for any }  \bt = (\bb, \bg) .\]
\noindent The theorem is proved in the appendix.   

\vskip 0.30in

An interpretation of the mNRI statistic is obtained by rewriting it as
\[ T_n(\hat{\bt}; \hat{\bt^0}, \hat{\pi}) = [2 \bar{y} (1-\bar{y})]^{-1} \frac{ [\bms(\hat{\bt}; \hat{\bt^0})]^T [\bmr(\hat{\bt^0})]}{[\bms(\hat{\bt}; \hat{\bt^0})]^T [\bms(\hat{\bt}; \hat{\bt^0})]} \]
where $\bmr(\hat{\bt^0}) = [r(\hat{\bb^0}^T \bmx_1), \ldots, r(\hat{\bb^0}^T \bmx_n)]$ is the base model score residual vector and $\bms(\hat{\bt}; \hat{\bt^0})$ is a sign vector with subject components $s_i(\hat{\bt};\hat{\bt^0}) = 2I(\hat{\bb}^T \bmx_i +\hat{\bg}^T \bmz_i - \hat{\bb^0}^T \bmx_i > 0 ) -1$. The mNRI is a function of the propensity of the event outcome $(\bar{y})$ and a regression coefficient representing the association between the direction of the risk score due to adding $\bmz$ and the event outcome after taking into account $\bmx$. This perspective is analogous to a partial residual plot, 
where a model covariate of interest $\bmz$ is replaced by a between model directional covariate $\bms(\hat{\bt}; \hat{\bt^0})$.

An alternative interpretation of the mNRI may be considered from the viewpoint of its limiting value
\[ \lim_{n \rightarrow \infty} T_n(\bt_0; \bt^0; \pi_0) = [2 \pi_0 (1-\pi_0)]^{-1} E_{X,Z} \left\{ h({\bb^0}^T \bmX) \left| G({\bb_0}^T \bmX + \bg_0^T \bmZ) - G({\bb^0}^T \bmX) \right| \right\} \]
where the weight $h({\bb^0}^T \bmX) $ stems from the base model score residual (4),
\[ h({\bb^0}^T \bmX) = \left[\frac{\partial G({\bb^0}^T \bmx)}{\partial ({\bb^0}^T \bmx)}\right] \left[G({\bb^0}^T \bmx)(1-G({\bb^0}^T \bmx))\right]^{-1} \]
\[   r({\bb^0}^T \bmX) = h({\bb^0}^T \bmX) [y_i - G({\bb^0}^T \bmX)] .\]
Thus, the population mNRI is a weighted $L_1$ distance measure between the nested event probabilities. An important special case occurs when $G$ is logistic and  
\[ \lim_{n \rightarrow \infty} T_n(\bt_0; \bt^0; \pi_0) = [2 \pi_0 (1-\pi_0)]^{-1} E_{X,Z} \big|G(\bb_0^T \bmX + \bg_0^T \bmZ) - G({\bb^0}^T \bmX)\big| , \]
which results in an unweighted $L_1$ distance measure. 
Here, the population mNRI is proportional to the mean absolute deviation (MAD) of the nested event probabilities. In addition to using the MAD as a summary measure, this result suggests that graphical insight into the mNRI may be obtained by plotting the base model event probability estimates by the expanded model event probability estimates.

\vskip 0.50in

\noindent {\bf 3. THE NRI FALSE POSITIVE RATE} 

\noindent Empirical research on the utility of the NRI has raised questions as to whether it has an unacceptably high false positive rate, signifying a larger than anticipated value when the new factors have no effect on the binary response (Kerr et al. 2014 and Pepe et al. 2014, 2015). As a practical matter, measures with high false positive rates lead to the introduction of irrelevant factors into the model development process. In this section, this issue is investigated, and a valid test procedure is developed, both in the case of a single sample and when independent training and test samples are included. 

Pencina et al (2008) state that under the null $\rho(\bt_0; \bt^0; \pi_0) = 0$, the asymptotic distribution of the estimated NRI in (2) is
\begin{equation} 
 n^{1/2} R_n(\hat{\bt}; \hat{\bt}^0; \hat{\pi}) \ \ \stackrel{D}{\rightarrow} \ \  
N \left[0, A \right] 
\end{equation} 
where accounting for the multiplication by $1/2$ to produce (3), the asymptotic variance is estimated as $\hat{A} = [(4n_1)^{-1} + (4n_0)^{-1}]$.
Further work by Pencina et al. (2011, 2012) modified the asymptotic variance calculation. 
In a series of simulation experiments, Kerr et al. (2014) and Pepe et al. (2014, 2015) evaluated the adequacy of this result,
using a conditional binormal model to produce nested logistic regression models. 
They found that on average, under the null, the NRI estimate was positive and that the type 1 error rate using the asymptotic normal reference distribution was as high as 0.63. Additional simulations that incorporated independent training and test datasets produced similar conclusions. 
Taken in total, these results represent a critical indictment against the test procedure in (5). A problem, recognized by these authors, and Demler et al. (2017), is that the asymptotic normal reference distribution is incorrect.  

Consider a smooth NRI 
\begin{equation}  R_n^{S}(\hat{\bt}; \hat{\bt}^0; \hat{\pi}) = \left[n \bar{y} (1-\bar{y})\right]^{-1} \sum_i \left[y_i - \bar{y} \right] \left[\Phi(\hat{\bb}^T \bmx_i +\hat{\bg}^T \bmz_i - \hat{\bb^0}^T \bmx_i ) - \frac{1}{2} \right] , \end{equation} 
where the extended indicator function, which is  discontinuous in $\bt$, is replaced by the continuous standard normal distribution function $\Phi(\cdot)$. A heuristic for this substitution is that when $\rho(\bt_0; \bt^0; \pi_0)=0$, as $n$ gets large, $ \ \hat{\bg} \stackrel{p}{\rightarrow} 0$, $\ \hat{\bb}-\hat{\bb^0} \stackrel{p}{\rightarrow} 0$ (Pepe et al. 2013), and therefore (Horowitz 1998)
\[ I(\hat{\bb}^T \bmx + \hat{\bg}^T \bmz - \hat{\bb^0}^T \bmx  > 0) \approx \Phi\left(\hat{\bb}^T \bmx +\hat{\bg}^T \bmz - \hat{\bb^0}^T \bmx  \right) \approx \frac{1}{2}.\] 
The purpose of this local smoothing is to facilitate the derivation of the asymptotic null reference distribution.
 
\vskip 0.30in 

\noindent {\bf Theorem 2. }  Assume the binary response regression models in (1) are properly specified and the covariate vectors $\bmx$ and $\bmz$ have dimension $p$ and $q$, respectively. If $\rho(\bt_0; \bt^0; \pi_0) = 0$, then
\[ n R_n^{S}(\hat{\bt}; \hat{\bt}^0; \hat{\pi}) \ \stackrel{D}{\rightarrow} \ [\bmN_q(0,V_1)]^T [\bmN_q(0,V_2)] + \frac{1}{2} \bmD_p^T \mathcal{I}_{\beta \beta}^{-1} \bmc_p. \] 

The first term is the inner product of two positively correlated, q-dimensional, mean zero normal random vectors, and the second term is bilinear, where $\bmD_p$ is a $p$ dimensional random vector with quadratic components, and $\bmc_p$ is a $p$ dimensional constant vector. This result demonstrates that the null distribution of the NRI is not normal and is not symmetric about zero, which explains the anomalous findings in Kerr et al. (2014) and Pepe et al. (2014, 2015). 

The reference distribution for the NRI test statistic $R_n^S(\hat{\bt}; \hat{\bt}^0; \hat{\pi})$ is complex and difficult to apply. In contrast, the mNRI test statistic 
\[  T_{n}^{S}(\hat{\bt}; \hat{\bt}^0; \hat{\pi}) = \left[n \bar{y} (1-\bar{y})\right]^{-1} \sum_i 
r(\hat{\bb^0}^T\bmx_i)  \left[\Phi(\hat{\bb}^T \bmx_i +\hat{\bg}^T \bmz_i - \hat{\bb^0}^T \bmx_i ) - \frac{1}{2} \right] \]
has a straightforward null reference distribution.

\vskip 0.30in

\noindent {\bf Theorem 3. } Assume the binary regression models in (1) are properly specified and the covariate vectors $\bmx$ and $\bmz$ have dimension $p$ and $q$, respectively.   If $\rho(\bt_0; \bt^0; \pi_0) = 0$, then
\[ n T_n^{S}(\hat{\bt}; \hat{\bt}^0; \hat{\pi}) \ \stackrel{D}{\rightarrow} \ k \chi_q^2 , \] 
where $k = \phi(0)[\pi_0 (1-\pi_0)]^{-1}$, $\ \phi(0)$ is the standard normal density function evaluated at 0, and $\chi_q^2$ is a chi-square random variable with $q$ degrees of freedom. A proof of this result is found in the appendix.

\vskip 0.30in

Theorems 2 and 3 reorient one's understanding of what constitutes meaningful NRI and mNRI statistics and Theorem 3 provides an uncomplicated  metric to test the mNRI  distance from zero. If the new clinical factors $(\bmz)$ are noise, then small positive values are simply random variation under the null, and only large positive values, as determined by the scaled chi-square reference distribution, are considered meaningful. 
A precursor to this result is found in Kerr et al. (2011). 

Theorem 3 covers the single sample case. Alternatively, the test statistic may be constructed from two independent data sets from the same population, where the regression coefficients are estimated from the training data $(\hat{\bt}, \hat{\bt}^0)$ and the test data $(\tilde{\bt}, \tilde{\bt}^0)$, and the data for the test statistic $(y_i, \bmx_i, \bmz_i)$ are drawn from the independent test data. Under these conditions, the reference distribution for the smooth mNRI test statistic 
\[  T_{n}^{S}(\hat{\bt}; \hat{\bt}^0,  \tilde{\bt}^0; \tilde{\pi}) = \left[n \bar{y} (1-\bar{y})\right]^{-1} \sum_i r(\tilde{\bb^0}^T\bmx_i) \left[\Phi(\hat{\bb}^T \bmx_i +\hat{\bg}^T \bmz_i - \hat{\bb^0}^T \bmx_i ) - \frac{1}{2} \right] . \]  
is provided in Theorem 4.

\vskip 0.30in

\noindent {\bf Theorem 4. }  Assume the binary regression models for the training and test data have the same specification and are given in (1), where the covariate vector $\bmx$ has dimension $p$ and the covariate vector $\bmz$ has dimension $q$.  If $\rho(\bt_0; \bt^0; \pi_0) = 0$,  
\[  n T_{n}^{s}(\hat{\bt}; \hat{\bt}^0,  \tilde{\bt}^0; \tilde{\pi})   \ \stackrel{D}{\rightarrow} \ \frac{k}{2} \sum_{j=1}^{2q} \lambda_j \chi^2_j  ,\]
\noindent  where $k$ is defined in Theorem 3, $\{ \chi_j^2\}$ are independent chi-square random variables each with one degree of freedom, and $\{\lambda_j\}$ represent eigenvalues determined from the product matrix $VC$ (Baldessari 1967), where
\[ 
V = \left( \begin{array}{cc}
\mbox{var}(\tilde{\bg}) & \ \ 0 \\ 
0 & \ \ \mbox{var}(\hat{\bg}) \end{array} \right) 
\ \ \ \ \ \ \ \ \ \ \ \ \ \ \ \ \ \ \ \ 
C = \left( \begin{array}{cc}
0 & \ \ D \\ 
D & \ \ 0 \end{array} \right).
\]
and $D = [\mbox{var}(\tilde{\bg})]^{-1} \ $. 
The details are provided in the appendix.

\vskip 0.30in 

A simulation study was performed to assess the false positive rate using the reference distributions in Theorems 3 and 4, and the normal reference distribution in (5). A conditional bivariate normal covariate distribution was used to generate nested logistic risk models. The conditioning variable was the event status with $\mbox{Pr}(Y=1) = \{ 0.25, 0.50, 0.75 \}$. The bivariate normal had a common variance-covariance matrix across event status, with  correlation parameters $0$ or $0.5$. The mean of $Z$ was $0$ for $Y=0$ or $Y=1$, and the mean of $X$ was set equal to $0$ for $Y=0$ and took on values $\{ 0.25, 0.50, 0.75, 1.0 \}$ for $Y=1$. Simulations with 200 and 500 observations per replicate were conducted.  Five thousand replicates were run for each simulation.    
Tables 1 and 2 compare the size estimates for the mNRI reference distributions in Theorems 3 and 4 with the NRI normal reference distribution. The nominal type 1 error in all simulations was 0.05. 

For the single sample simulations in Table 1, using Theorem 3, the average type 1 error was 0.048 (n=200) and 0.050 (n=500). In contrast, applying the normal reference distribution in (5), produced average type 1 errors equal to 0.079 (n=200) and 0.129 (n=500). Similar results were found for the independent training-test sample simulations in Table 2. From Theorem 4, the average type 1 error was 0.051 (n=200) and 0.052 (n=500), whereas when using the normal reference distribution it was 0.079 (n=200) and 0.124 (n=500).  These simulation results confirm that the modified NRI test statistics, with their associated reference distributions, are valid test procedures, and they confirm the poor operating characteristics of the asymptotic normal reference distribution, with divergence increasing with sample size.

\vskip 0.50in

\noindent {\bf 5. PROSTATE CANCER DATA} 

Patients with metastatic prostate cancer are by definition high risk. Nevertheless, there is significant variability in the survival times of these patients (Sayegh, Swami, and Agarwal, 2021). Given this heterogeneity, there is a pressing need to identify new biomarkers that can accurately assess patient risk. Historically, the use of prostate specific antigen (PSA) and other blood based biomarkers have produced risk models with only moderate calibration and discrimination in the metastatic prostate cancer setting (Gafita et al. 2021). As a result, exploring informative new biomarkers continues, with a recent focus around circulating tumor cells and serum testosterone (Cieslikowski et al. 2021; Ryan et al. 2019).

An application of the net reclassification improvement (NRI), based on the addition of circulating tumor cells and serum testosterone, was undertaken for metastatic prostate cancer patients treated on the control arm of a multicenter phase 3 randomized clinical trial (Saad et al. 2015). The control arm of the randomized trial,  patients treated with steroids alone, is useful to assess the added prognostic utility of new biomarkers, because it approximates the natural history of the disease.   

Four hundred and eighteen patients with a complete set of biomarkers and sufficient follow-up were used in the analysis. The binary endpoint was survival 24 months after the start of treatment. In this cohort, forty seven percent of the patients survived longer than two years. In addition to circulating tumor cells and serum testosterone, traditional biomarkers for metastatic prostate cancer were incorporated into the risk model. The complete set of eight biomarkers included in the analysis were: albumin, alkaline phosphatase, circulating tumor cells, Gleason score, hemoglobin, lactate dehydrogenase, prostate specific antigen, and serum testosterone. Nested logistic regression models were fit for the binary 24 month survival endpoint; the expanded model incorporated all eight biomarkers and the base model represented a subset of seven biomarkers. All biomarkers except Gleason score were continuous. To create greater flexibility in the models, a restricted cubic spline with four knots was fit to each continuous biomarker. Gleason score, an ordinal variable ranging from 2-10, representing tumor complexity as determined by pathology, and was dichotomized as 1-7 and 8-10.   

Table 3 summarizes the results of the NRI, mNRI, and the p-values generated from their respective test procedures described in Section 3.  For the logistic models, the mNRI equates to a scaled mean absolute difference (MAD) between the estimated event probabilities 
\[ [2n\bar{y}(1-\bar{y})]^{-1} \sum_i |G(\hat{\bb}^T \bmx_i + \hat{\bg}^T \bmz_i) - G(\hat{\bb^0}^T \bmx_i) |  .\]
For the prostate data, the observed proportion of events was 0.47, and so the $\mbox{mNRI} \approx 2 \times \mbox{MAD}$.
  
With the addition of serum testosterone, the mean absolute distance was only 0.022, and using the smooth mNRI, a test of whether the population NRI differed from zero generated a p-value equal to 0.490. Figure 1 provides corroborating evidence that adding serum testosterone does not meaningfully change the predicted event probabilities. An application of the NRI with a normal reference distribution (5), however, produced a p-value equal to 0.046, which mirrors the high false positive rate for the NRI found in the simulations.  When the circulating tumor cell (CTC) biomarker was added to the risk model, the mean absolute difference between the estimated event probabilities was large and equal to 0.095, with an attending p-value less than 0.001. The addition of circulating tumor cells had a marked effect on the predicted probability of death within 24 months. This result is confirmed visually in Figure 2, where the estimated event probabilities change significantly from the base model to the expanded model due to the addition of CTC. Thus, the addition of CTC but not serum testosterone would consequentially change the predicted probabilities of surviving greater than 24 months. Furthermore, for other single variable deletions, only the addition of alkaline phosphatase and hemoglobin appreciably change the expanded model probabilities.

\vskip 0.50in

\noindent {\bf 6. DISCUSSION} 

\noindent The net reclassification improvement (NRI) statistic is a measure of change for a model based risk score due to the addition of new factors. 
Although the NRI is frequently applied, identified weaknesses of the statistic include that it is not a proper scoring function (or proper change score) and it does not produce a valid test procedure. A modification of this statistic (mNRI) corrects these deficiencies. The mNRI can be interpreted as a measure of association between the directional change in the risk score and the base model score residual. In the special but frequently applied case of logistic regression, an asymptotic analysis demonstrates that the mNRI is proportional to a mean absolute deviation measure, putting the mNRI on an easily interpretable difference in probability scale. 

There remain, however, some concerns with the NRI that are not resolved through the mNRI (Kerr et al. 2014). The mNRI does not include risk thresholds for the purpose of intervention strategies, and therefore does not include the costs and benefits of a risk threshold guided intervention. As a result, its application should be directed to the model development stage. On this topic, there has been significant discussion surrounding the utility of the NRI, and even with the modification proposed here, the debate will almost surely continue.  The contribution of this work is to put the statistic on a stronger statistical foundation and to clear away some of the arguments that obscure its properties, perhaps shedding more light and less heat on this measure.

\vskip 0.50in

\noindent {\bf ACKNOWLEDGEMENTS.  }
\newline This work was supported by NIH Grants R01CA207220 and P30CA008748.

\newpage

\noindent{\bf REFERENCES} 
\newcounter{refr}
\noindent
\begin{list}{}{\usecounter{refr} \setlength{\itemindent}{-10pt}
\setlength{\itemindent}{-10pt}
\setlength{\itemsep}{-2pt}}

\bigskip 

\item
Baldessari, B. (1967), "The Distribution of a Quadratic Form of Normal Random Variables," \emph{Annals of Mathematical Statistics}, 38, 1700-1704.

\item
Bickel, P. J., Klaassen, C. A. J., Ritov, Y., and Wellner, J. A. (1993), \emph{Efficient and Adaptive Estimation for Semiparametric Models}, The Johns Hopkins University Press.

\item
Cieslikowski, W. A., Antczak, A., Nowicki, M.,  Zabel, M., Budna-Tukan, J. (2021), "Clinical Relevance of Circulating Tumor Cells in Prostate Cancer Management," \emph{Biomedicines}, 9, 1179.

%\item
%Cox, D. R. and Hinkley, D. V. (1974), \emph{Theoretical Statistics}. Chapman and Hall.

\item
Demler, O. V., Pencina, M. J., Cook, N. R., and D'Agostino Sr, R. B. (2017),
"Asymptotic distribution of $\Delta$AUC, NRIs, and IDI based on theory of U-statistics, "\emph{Statistics in Medicine}, 36, 3334-3360.

\item
Gafita, A., Calais, J., Grogan, T. R., Hadaschik, B., Wang, H., Weber, M., Sandhu, S., Kratochwil, C., Esfandiari, R., Tauber, R., Zeldin, A., Rathke, H., Armstrong, W. R., Robertson, A., Thin, P., D'Alessandria, C., Rettig, M. B., Delpassand, E. S., Haberkorn, U., Elashoff, D., Herrmann, K., Czernin, J., Hofman, M. S., Fendler, W. P., Eiber, M. (2021), "Nomograms to predict outcomes after 177 Lu-PSMA therapy in men with metastatic castration-resistant prostate cancer: an international, multicentre, retrospective study," \emph{Lancet Oncology}, 22, 1115–25.

\item
Gerds, T. A. and Kattan, M. W. (2021), \emph{Medical Risk Prediction Models With Ties to Machine Learning}. CRC Press.

\item
Gneiting, T. and Raftery, A. E. (2007), "Strictly proper scoring rules, prediction, and estimation," \emph{Journal of The American Statistical Association}, 102, 359-378.

\item
Horowitz, J. L. (1998), \emph{Semiparametric Methods in Econometrics}. Springer-Verlag. 

\item
Kerr, K. F., McClelland, R. L., Brown, E. R., and Lumley, T. (2011), "Evaluating the Incremental Value of New Biomarkers With Integrated Discrimination Improvement," {\emph American Journal of Epidemiology}, 174, 364-374. 

\item
Kerr, K. F., Wang, Z., Janes, H., McClelland, R. L., Psaty, B. M., and Pepe, M. S. (2014), "Net reclassification indices for evaluating risk-prediction instruments: A critical review," \emph{Epidemiology}, 25, 114-121.

\item
Manski, C. F. (1985), "Semiparametric analysis of discrete response: Asymptotic properties of the maximum score estimator," \emph{Journal of Econometrics}, 27, 313-333.

\item
Pencina, M. J., D'Agostino Sr, R. B., D'Agostino Jr, R. D., and Vasan, R.  (2008), "Evaluating the added predictive ability of a new marker: From area under the ROC curve to reclassification and beyond," \emph{Statistics in Medicine}, 27, 157-172.

\item
Pencina, M. J., D'Agostino Sr, R. B., and Steyerberg, E. W. (2011), "Extensions of net reclassification improvement calculations to measure usefulness of new biomarkers," \emph{Statistics in Medicine}, 30, 11-21. 

\item
Pencina, M. J., D'Agostino Sr, R. B., and Demler O. V. (2012), "Novel metrics for evaluating improvement in discrimination: net reclassification and integrated discrimination improvement for normal variables and nested models," \emph{Statistics in Medicine}, 31, 101-113. 

\item 
Pepe, M. S., Fan, J., Feng, Z., Gerds, T., and Hilden, J. (2015), "The net reclassification index (NRI): A misleading measure of prediction improvement even with independent test data sets," \emph{Statistics in Biosciences}, 7, 282-295.

\item
Pepe, M. S., Janes, H., and Li, C. I. (2014), Net risk reclassification p values: Valid or misleading? \emph{Journal of the National Cancer Institute}, 106, 1-6.

\item
Pepe, M. S., Kerr, K. F., Longton, G., and Wang, Z. (2013), "Testing for improvement in prediction model performance," \emph{Statistics in Medicine}, 32, 1467-1482.

\item
Ryan, C. J., Dutta, S., Kelly, W. K., Russell, C., Small, E. J., Morris, M. J., Taplin, M. E., Halabi, S. (2020), "Androgen Decline and Survival During Docetaxel Therapy in Metastatic Castration Resistant Prostate Cancer (mCRPC)," \emph{Prostate Cancer and Prostatic Disease}, 23, 66-73.

\item
Saad, F., Fizazi, K., Jinga, V., Efstathiou, E., Fong, P. C., Hart, L. L., Jones, R., McDermott, R., Wirth, M., Suzuki, K., MacLean, D. B., Wang, L., Akaza, H., Nelson, J., Scher, H. I., Dreicer, R., Webb, I. J., de Wit, R. ELM-PC 4 investigators. (2015), "Orteronel plus prednisone in patients with chemotherapy naive metastatic castration-resistant prostate cancer (ELM-PC 4): a double-blind, multicentre, phase 3, randomised, placebo-controlled trial," \emph{Lancet Oncology}, 16, 338–348.

\item
Sayegh, N., Swami, U., and Agarwal, N. (2021), "Recent Advances in the Management of Metastatic Prostate Cancer," \emph{JCO Oncology Practice}, 18, 45-55.

\item
Tsiatis, A. A. (2006), \emph{Semiparametric Theory and Missing Data}. Springer-Verlag.

\end{list}

\newpage

\renewcommand{\theequation}{A.\arabic{equation}}  
\setcounter{equation}{0}  

\noindent {\sc Appendix: Proof of theorems} 

\vskip 0.30in

\noindent The following conditions and notation will be used in the appendix.

\vskip 0.15in

\noindent (C1)  The set of binary response nested models
\begin{align*} 
\begin{split}
\mbox{Pr}(Y=1) & = G(\beta^{\bullet}) \\
\mbox{Pr}(Y=1 |\bmx) & = G({\bb^0}^T \bmx) \\
\mbox{Pr}(Y=1 |\bmx, \bmz) & = G(\bb_0^T \bmx + \bg_0^T \bmz) 
\end{split}
\end{align*}
specify the relationship between the $p$-dimensional existing factors $\bmx$, the $q$-dimensional new factors $\bmz$, and the binary event outcome $y$. The model with no covariates is the constant model, $\bmx$ alone is the base model and $(\bmx, \bmz)$ is the expanded model. The inverse link function $G$ is known. Throughout this work, random variables are represented with upper case, their observed copies are written in lower case, and vectors are indicated in bold.

\bigskip 

\noindent (C2) The log-likelihood used to estimate the regression coefficients is  
\[ L(\bb, \bg) = \sum_i \left[y_i \log G(\bb^T \bmx_i + \bg^T \bmz_i) + (1-y_i) \log (1-G(\bb^T \bmx_i + \bg^T \bmz_i)) \right] ,\]
where $\{(y_i,\bmx_i, \bmz_i)\}, i=1, \ldots, n$ are independent identically distributed copies of $(Y, \bmX, \bmZ)$. For $\bt = (\bb, \bg)$, the expanded model maximum likelihood estimate is denoted by $\hat{\bt} = (\hat{\bb}, \hat{\bg})$, and the two sets of restricted maximum likelihood estimates are $\hat{\bt}^0 = (\hat{\bb}^0, \bmzero)$ for the base model, and $\hat{\pi} = G(\hat{\bb}^{\bullet})$, which is equal to the mean number of events $\bar{y}$, for the constant model. 

\bigskip

\noindent (C3) The score vector, observed information matrix, and expected information matrix for $L(\bt)$ are partitioned as
\[  \frac{\partial L(\bt)}{\partial \bt}  =  \left( \begin{array}{c}
U_{\beta}  \\ U_{\gamma}  \end{array} \right) ; \ \ \ \ \ 
\frac{\partial^2 L(\bt)}{\partial \bt \partial \bt^T} =  \left( \begin{array}{cc}
U_{\beta \beta} &  U_{\beta \gamma} \\ U_{\gamma \beta} &  U_{\gamma \gamma} \end{array} \right) ; \ \ \ \ \  - \mbox{E}\left[n^{-1} U_{\theta \theta}\right]   
  =  \left( \begin{array}{cc}
\mathcal{I}_{\beta \beta} &  \mathcal{I}_{\beta \gamma} \\ \mathcal{I}_{\gamma \beta} & \mathcal{I}_{\gamma \gamma} \end{array} \right) 
 \] 

\bigskip

\noindent (C4)  The likelihood parameterization $L(\bn)$ will be utilized, where $\eta_i = \bb^T \bmx_i + \bg^T \bmz_i$ is the risk score and the corresponding score residual $r(\eta_i)$ is
\[ \frac{\partial L(\bn)}{\partial \eta_i} = \left(\frac{dG(\eta_i)}{d\eta_i}\right) \left[G(\eta_i)(1-G(\eta_i))\right]^{-1} \left[y_i - G(\eta_i)\right] ,\]
which will be useful to rewrite as
\[  r(\eta_i) = h(\eta_i) [y_i - G(\eta_i)] .\]
 
\vskip 0.50in

\noindent {\it Proof of Theorem 1:  The modified NRI (mNRI) is a proper change score}

\vskip 0.20in

\noindent For a single random variable, the modified NRI with the base and constant model parameters evaluated at their true value is
\[ T_1(\bt; \bt^0; \pi_0) = \left[ \pi_0 (1-\pi_0) \right]^{-1} r({\bb^0}^T\bmX)  \left[I(\bb^T \bmX +\bg^T \bmZ - {\bb^0}^T \bmX > 0 ) - \frac{1}{2} \right] .\]

\noindent Its expected value is equal to 
\begin{eqnarray*}
\lefteqn{ \mbox{E}_{X,Z} \left\{ \left[ \pi_0 (1-\pi_0) \right]^{-1} h({\bb^0}^T\bmX)  \right. \times }  \\
 & & \left. \left[G(\bb_0^T \bmX + \bg_0^T \bmZ) - G({\bb^0}^T \bmX) \right]  
\left[I(\bb^T \bmX + \bg^T \bmZ - {\bb^0}^T \bmX > 0)  - \frac{1}{2}  \right]   \right\} 
\end{eqnarray*}
where $h({\bb^0}^T\bmX)$ is a component of the score residual in (C4) evaluated under the base model.

\vskip 0.25in

\noindent To show $\mbox{E}[T_1(\bt; \bt^0; \pi_0)]$ is maximized at $\bt = \bt_0$, and therefore the modified NRI is a proper change score, consider
\begin{eqnarray*}
\lefteqn{\mbox{E}[T_1(\bt_0; \bt^0; \pi_0) - T_1(\bt; \bt^0; \pi_0)] = } \\
 & & \mbox{E}_{X,Z} \left\{ \left[ \pi_0 (1-\pi_0) \right]^{-1} h({\bb^0}^T\bmX) \left[G({\bb_0}^T \bmX + {\bg_0}^T \bmZ) - G({\bb^0}^T \bmX) \right] \right. \ \times \\
 & & \left. \left[I({\bb_0}^T \bmX + {\bg_0}^T \bmZ - {\bb^0}^T \bmX > 0) - I(\bb^T \bmX + \bg^T \bmZ - {\bb^0}^T \bmX > 0) \right] \right\} 
\end{eqnarray*}

\bigskip

\noindent This expectation is evaluated under two cases:

\bigskip

\noindent {\it Case (i): } $\bb_0^T \bmX + \bg_0^T \bmZ \ge {\bb^0}^T \bmX$

\noindent The first term in square brackets, $G({\bb_0}^T \bmX + {\bg_0}^T \bmZ) - G({\bb^0}^T \bmX)$, is non-negative due to the monotonicity of $G$, and the second term in square brackets, the difference in indicator functions, is either $0$ or $1$. Therefore, since the weight function $h({\bb^0}^T\bmX)$ is positive, the expectation is non-negative for any $\bt = (\bb, \bg)$.

\bigskip 

\noindent {\it Case (ii): } $\bb_0^T \bmX + \bg_0^T \bmZ < {\bb^0}^T \bmX$. 

\noindent Under this constraint, the first term in square brackets is negative and the second term in square brackets is either $0$ or $-1$. It follows that the expectation is again non-negative for any $\bt = (\bb, \bg)$.

\vskip 0.20in

\noindent Combining these two cases, $\mbox{E}[T_1(\bt; \bt^0; \pi_0)]$ is maximized at $\bt = \bt_0$ and therefore, the modified NRI is a proper change score.

\vskip 0.45in

\noindent {\bf Theorem 2. }  Assume the covariate vectors $\bmx$ and $\bmz$ have dimension $p$ and $q$, respectively.  If $\rho(\bt_0; \bt^0; \pi_0) = 0$, then the smooth NRI test statistic 
\[ n R_n^{S}(\hat{\bt}; \hat{\bt}^0; \hat{\pi}) \ \stackrel{D}{\rightarrow} \ [\bmN_q(0,V_1)]^T [\bmN_q(0,V_2)] + \frac{1}{2} \bmD_p^T \mathcal{I}_{\beta \beta}^{-1} \bmc_p. \] 

\noindent The first term is the inner product of two positively correlated, q-dimensional, mean zero normal random vectors, and the second term is bilinear, where $\bmD_p$ is a $p$ dimensional random vector with quadratic components, and $\bmc_p$ is a $p$ dimensional constant vector.  

\vskip 0.30in

\noindent {\it Proof of Theorem 2:}

\noindent The smooth NRI test statistic is
\[  R_n^{S}(\hat{\bt}; \hat{\bt^0}; \hat{\pi}) = \bigg[n \bar{y} (1-\bar{y})\bigg]^{-1} \sum_i \left[y_i - \bar{y} \right] \left[\Phi(\hat{\bb}^T \bmx_i +\hat{\bg}^T \bmz_i - \hat{\bb^0}^T \bmx_i ) - \frac{1}{2} , \right] , 
\]
where $\Phi(\cdot)$ is the standard normal distribution function.

\bigskip

\noindent To determine its null reference distribution, Pepe et al. (2013) demonstrate that for correctly specified nested models (C1), $\rho(\bt_0; \bt^0; \pi_0) = 0 \ $ {\it iff} $ \ \bg_0 = 0$. This allows consideration of a second order Taylor expansion of $ R_n^{S}(\hat{\bt}; \hat{\bt}^0; \hat{\pi})$  around $\hat{\bt} = \hat{\bt}^0$, 
\begin{equation} 
n R_n^{S}(\hat{\bt}; \hat{\bt^0}; \hat{\pi}) = \left[\frac{\phi(0)}{ \bar{y} (1-\bar{y})}\right] \sum_i \left[ (\hat{\bb} - {\hat{\bb}^0})^T \bmx_i + (\hat{\bg} - {\hat{\bg}}^0)^T \bmz_i \right] \left[y_i - \bar{y}\right]  + o_p(1), 
\end{equation}
where $\phi(0)$ represents the standard normal density function evaluated at 0, and since its derivative evaluated at zero, $\phi'(0) = 0$, each element of the matrix in the quadratic term of the expansion is equal to 0.

\bigskip

\noindent To further simplify, note that
\begin{equation} n^{1/2} (\hat{\bb} - {\hat{\bb}}^0) = -\mathcal{I}^{-1}_{\beta \beta} \mathcal{I}_{\beta \gamma} [n^{1/2} (\hat{\bg} - \bg_0)] + (4n)^{-1/2} \mathcal{I}^{-1}_{\beta \beta} \bmd(\hat{\bt}; \hat{\bt}^0) + o_p(n^{-1/2}) 
\end{equation}
which follows from a second order Taylor series approximation of the score statistic (C3), $U_{\beta}(\bt)$ around $\hat{\bt} = \hat{\bt}^0$, with 
\[ \bmd(\hat{\bt}; \hat{\bt}^0) = \left[ \begin{array}{c}
n^{1/2}(\hat{\bt} - \hat{\bt}^0)^T [n^{-1} H^{(1)}(\hat{\bt}^0)] n^{1/2}(\hat{\bt} - \hat{\bt}^0) \\ \vdots \\ n^{1/2}(\hat{\bt} - \hat{\bt}^0)^T [n^{-1} H^{(p)}(\hat{\bt}^0)] n^{1/2}(\hat{\bt} - \hat{\bt}^0)  \end{array} \right]  \ \  \mbox{ and   } \ \ H^{(j)}(\bt) = \frac{\partial^2 U_{\beta_j}(\bt)}{\partial \bt \partial \bt^T} \ . \]

\bigskip

\noindent Substituting (A.2) into (A.1), 
\begin{eqnarray}
\lefteqn{n R_n^{S}(\hat{\bt}; \hat{\bt}^0; \hat{\pi}) = \left[\frac{\phi(0)}{\bar{y} (1-\bar{y})}\right]  \times } \\
  &  \left\{\left[n^{1/2}(\hat{\bg} - \bg_0)\right]^T \left[n^{-1/2} \sum_i\left(\bmz_i - \mathcal{I}_{\gamma \beta} \mathcal{I}^{-1}_{\beta \beta} \bmx_i\right)  
(y_i - \bar{y}) \right] \right. +  \nonumber \\ 
 &  \left. \frac{1}{2} \left[ \bmd(\hat{\bt}; \hat{\bt}^0) \right]^T \mathcal{I}^{-1}_{\beta \beta} \left[ n^{-1} \sum_i \bmx_i (y_i-\bar{y}) \right] \right\}   + o_p(1) .\nonumber
	\end{eqnarray}

\bigskip

\noindent To obtain the result in Theorem 2, consider the elements in (A.3), 
\[ \frac{\phi(0)}{ \bar{y} (1-\bar{y})} \ \stackrel{p}{\rightarrow} \ \frac{\phi(0)}{\pi_0 (1-\pi_0)} \]
\[ n^{1/2}(\hat{\bg} -\bg_0) \ \stackrel{D}{\rightarrow} \  \bmN_q(0,V_\gamma) \]
\[ n^{-1} \sum_i \bmx_i (y_i - \bar{y}) \ \stackrel{p}{\rightarrow} \ \bmc_p \]
\[ \bmd(\hat{\bt}; \hat{\bt}^0) \ \stackrel{D}{\rightarrow} \ \bmD_p \]

\noindent The remaining element is
\[ n^{-1/2} \sum_i\left( \bmz_i   - \mathcal{I}_{\gamma \beta} \mathcal{I}^{-1}_{\beta \beta} \bmx_i  \right)  \left( y_i - \bar{y} \right) . \]

\noindent First, under the null
\[ n^{-1} \sum_i\left(\bmz_i  - \mathcal{I}_{\gamma \beta} \mathcal{I}^{-1}_{\beta \beta} \bmx_i \right)  (y_i - \bar{y}) \ \stackrel{p}{\rightarrow} \ E_{X,Z}\left\{ (\bmZ  -  \mathcal{I}_{\gamma \beta} \mathcal{I}^{-1}_{\beta \beta} \bmX )(G({\bb^0}^T \bmX) - \pi_0) \right\} \]

\noindent which is rewritten as
\begin{equation} E_{X,Z}\left\{ (\bmZ_{*}  -  \mathcal{I}_{\gamma \beta} \mathcal{I}^{-1}_{\beta \beta} \bmX_{*})(W_X^{-1/2}[G({\bb^0}^T \bmX) - \pi_0 ] ) \right\} \end{equation}
\noindent where $\bmZ_{*} = \bmZ W_X^{1/2}, \ \bmX_{*} = \bmX W_X^{1/2}, \mbox{ and } \
W_X = \mbox{var}[r({\bb^0}^T \bmX) | \bmX]$.

\bigskip

\noindent The motivation for the weight $W_X$ comes from the Bernoulli loglikelihood (C2, C3)
\[ \mathcal{I}_{\beta \beta} = E[ \bmX_{*} \bmX_{*}^T] \ \ \ \ \ \mathcal{I}_{\gamma \beta} = E[\bmZ_{*} \bmX_{*}^T] \]
and the recognition that 
\[ E_{X,Z}\left\{ (\bmZ_{*}  -  \mathcal{I}_{\gamma \beta} \mathcal{I}^{-1}_{\beta \beta} \bmX_{*}) \bmX_{*}^T \right\} = 0 ,\]
a $q \times p$ matrix of zeros. 

\noindent Therefore by projection theory (Tsiatis 2006),
\[ E[\bmZ_{*} | \bmX_{*}] = \mathcal{I}_{\gamma \beta} \mathcal{I}^{-1}_{\beta \beta} \bmX_{*} \]
and so the expectation in (A.4) is equal to zero.

\bigskip

\noindent It now follows from the central limit theorem, 
\[ n^{-1/2} \sum_i\left(\bmz_i  - \mathcal{I}_{\gamma \beta} \mathcal{I}^{-1}_{\beta \beta} \bmx_i \right)  (y_i - \bar{y})  \ \stackrel{D}{\rightarrow} \  \bmN_q(\bmzero,V_2) .\]

\vskip 0.20in

\noindent Theorem 2 is the result of Slutsky's theorem applied to the elements in (A.3).

\vskip 0.45in

\noindent {\bf Theorem 3. } Assume the binary regression models in (C1) are properly specified and the covariate vectors $\bmx$ and $\bmz$ have dimension $p$ and $q$, respectively.   If $\rho(\bt_0; \bt^0; \pi_0) = 0$, then
\[ n T_n^{S}(\hat{\bt}; \hat{\bt}^0; \hat{\pi}) \ \stackrel{D}{\rightarrow} \ k \chi_q^2 , \] 
where $k = \phi(0)[\pi_0 (1-\pi_0)]^{-1}$, $\ \phi(0)$ is the standard normal density function evaluated at 0, and $\chi_q^2$ is a chi-square random variable with $q$ degrees of freedom.

\vskip 0.30in

\noindent {\it Proof of Theorem 3:} 

\noindent The mNRI test statistic is,
\[
 n T_{n}^{S}(\hat{\bt}; \hat{\bt^0}; \hat{\pi}) = \left[\bar{y} (1-\bar{y})\right]^{-1} \sum_i r(\hat{\bb^0}^T \bmx_i) \left[\Phi(\hat{\bb}^T \bmx_i +\hat{\bg}^T \bmz_i - \hat{\bb^0}^T \bmx_i ) - \frac{1}{2} \right], \]  
where $r(\cdot)$ is the score residual defined in (C4). 

\noindent A second order Taylor expansion of $T_{n}^{S}$ around $\hat{\bt} = \hat{\bt}^0$ results in
\newline $n T_n^{S}(\hat{\bt}; \hat{\bt^0}; \hat{\pi}) = $
\begin{equation*} \frac{\phi(0)}{\bar{y} (1-\bar{y})} \left[ n^{1/2}(\hat{\bg} - \bg_0) \right]^T \left[n^{-1/2} \sum_i r(\hat{\bb^0}^T \bmx_i) \left(\bmz_i - \mathcal{I}_{\gamma \beta} \mathcal{I}^{-1}_{\beta \beta} \bmx_i\right)  \right] + o_p(1) .
\end{equation*}

\bigskip
\noindent This approximation may be further simplified through the recognition that  
\newline $\sum_i r(\hat{\bb^0}^T \bmx_i) \left(\bmz_i - \mathcal{I}_{\gamma \beta} \mathcal{I}^{-1}_{\beta \beta} \bmx_i\right)$
is the efficient score statistic for estimating $\bg$ in the presence of $\bb$ and evaluated under the constraint $\bg=0$.  It follows that (Bickel, Klassen, Ritov, and Wellner, 1993) 
\[ n^{-1/2} \sum_i r(\hat{\bb^0}^T \bmx_i) \left(\bmz_i - \mathcal{I}_{\gamma \beta} \mathcal{I}^{-1}_{\beta \beta} \bmx_i\right)  =  [\mathcal{I}^{\gamma \gamma}]^{-1}[n^{1/2}(\hat{\bg} - \bg_0)] + o_p(1) ,\]
and therefore,
\[ n T_n^{S}(\hat{\bt}; \hat{\bt}^0; \hat{\pi}) = \frac{\phi(0)}{\pi_0 (1-\pi_0)} \left[ n^{1/2}(\hat{\bg} - \bg_0) \right]^T [\mathcal{I}^{\gamma \gamma}]^{-1}[n^{1/2}(\hat{\bg} - \bg_0)] + o_p(1) .\]
That is, 
\[ \mbox{Pr}\left(n T_n^{S}(\hat{\bt}; \hat{\bt}^0; \hat{\pi}) \le u \right) = \mbox{Pr}\left( k \chi_q^2 \le u \right) 
\] 
where $k = \phi(0)[\pi_0 (1-\pi_0)]^{-1}$ and $\chi_q^2$ is a chi-square random variable with $q$ degrees of freedom.

\vskip 0.45in

\noindent {\bf Theorem 4. }  Assume the binary regression models for the training and test data have the same specification and are given in (C1), where the covariate vector $\bmx$ has dimension $p$ and the covariate vector $\bmz$ has dimension $q$. Denote the estimated regression coefficients from the training data by $(\hat{\bt}, \hat{\bt}^0, \hat{\pi})$, the coefficients from the test data by $(\tilde{\bt}, \tilde{\bt}^0, \tilde{\pi})$, and the data  $(y_i, \bmx_i, \bmz_i)$ are drawn from the test sample. If $\rho(\bt_0; \bt^0; \pi_0) = 0$,  
\[  n T_{n}^{S}(\hat{\bt}; \hat{\bt}^0,  \tilde{\bt}^0; \tilde{\pi})   \ \stackrel{D}{\rightarrow} \ \frac{k}{2} \sum_{j=1}^{2q} \lambda_j \chi^2_j  ,\]
\noindent  where $k$ is defined in Theorem 3, $\{ \chi_j^2\}$ are independent chi-square random variables each with one degree of freedom, and $\{\lambda_j\}$ represent eigenvalues determined from the product matrix $VC$, where
\[ 
V = \left( \begin{array}{cc}
\mbox{var}(\tilde{\bg}) & \ \ 0 \\ 
0 & \ \ \mbox{var}(\hat{\bg}) \end{array} \right) 
\ \ \ \ \ \ \ \ \ \ \ \ \ \ \ \ \ \ \ \ 
C = \left( \begin{array}{cc}
0 & \ \ D \\ 
D & \ \ 0 \end{array} \right).
\]
and $D = [\mbox{var}(\tilde{\bg})]^{-1} \ $. 

\vskip 0.30in

\noindent {\it Proof of Theorem 4:} 

\vskip 0.20in

\noindent The test statistic for the NRI derived from training and test data are 
\[ n T_{n}^{S}(\hat{\bt}; \hat{\bt}^0,  \tilde{\bt}^0; \tilde{\pi}) = \left[\bar{y} (1-\bar{y})\right]^{-1} \sum_i r(\tilde{\bb^0}^T \bmx_i)  \left[\Phi(\hat{\bb}^T \bmx_i + \hat{\bg}^T \bmz_i - \hat{\bb^0}^T \bmx_i ) - \frac{1}{2} \right] . \]

\noindent Employing the arguments provided in the proof of Theorem 3, the smooth mNRI may be asymptotically approximated by 
\[ n T_{n}^{S}(\hat{\bt}; \hat{\bt}^0,  \tilde{\bt}^0; \tilde{\pi}) = \left[\frac{\phi(0)}{\pi_0 (1-\pi_0)}\right] [n^{1/2}(\hat{\bg} - \bg_0)]^T [I^{\gamma \gamma}]^{-1} [n^{1/2}(\tilde{\bg} - \bg_0)] + o_p(1) . \] 

\noindent The test statistic $T_n^{S}$ is bilinear, due to the different coefficient estimates $(\hat{\bg}, \tilde{\bg})$ from the training and test data. This statistic may be transformed to the quadratic 
\[ \frac{k}{2} \ \left[
\begin{array}{c}
\left(\hat{\bg} -\bg_0 \right) \\
 \left(\tilde{\bg} -\bg_0 \right)
\end{array}
\right]^T  
\left( 
\begin{array}{cc}
0 & D \\
D & 0 
\end{array}
\right)
\left[
\begin{array}{c}
\left(\hat{\bg} -\bg_0 \right) \\
 \left(\tilde{\bg} -\bg_0 \right)
\end{array}
\right] .
\]

\noindent It follows from Baldessari (1967) that as $n \rightarrow \infty$,
\[ \mbox{Pr}\left( n T_{n}^{S}(\hat{\bt}; \hat{\bt}^0,  \tilde{\bt}^0; \tilde{\pi}) \le t \right) = \mbox{Pr}\left(\frac{k}{2}\sum_{j=1}^{2q} \lambda_j \chi^2_j \le t\right) .\]

\newpage

\begin{center}
{\bf TABLE 1. } Type 1 error for the NRI and the modified NRI
\newline test procedures using a single sample

\vskip 0.25in
\renewcommand{\arraystretch}{0.65}

\begin{tabular}{|c|c|c|c|c|c|c|c|}
\multicolumn{3}{c}{ } &
\multicolumn{2}{c}{$\rho = 0$} &
\multicolumn{1}{c}{ } &
\multicolumn{2}{c}{$\rho = 0.5$} \\
\hline

$n$ & $ \pi_0$ &$\mu_X$&  mNRI &  NRI   &  & mNRI   & NRI      \\ 
		&					 &			 &  test &  test  &  & test   & test      \\ \hline
200 & 	0.25 & 	0.25 & 	0.0494 & 0.0468 &  & 0.0496 & 0.0504   \\
    & 	     & 	0.50 & 	0.0504 & 0.0578 &  & 0.0466 & 0.0582   \\
    & 	     & 	0.75 & 	0.0484 & 0.0776 &  & 0.0470 & 0.0724   \\
    &  	     & 	1.00 & 	0.0452 & 0.1046 &  & 0.0494 & 0.1034   \\ \hline
    & 	0.50 & 	0.25 & 	0.0508 & 0.0574 &  & 0.0516 & 0.0678   \\
    & 	     & 	0.50 & 	0.0488 & 0.0820 &  & 0.0546 & 0.0804   \\
    & 	     & 	0.75 & 	0.0538 & 0.1028 &  & 0.0500 & 0.1008   \\
    & 	     & 	1.00 & 	0.0510 & 0.1242 &  & 0.0444 & 0.1276   \\ \hline
    & 	0.75 & 	0.25 & 	0.0454 & 0.0466 &  & 0.0444 & 0.0466   \\ 
    & 	     & 	0.50 & 	0.0432 & 0.0588 &  & 0.0462 & 0.0568   \\
    & 	     & 	0.75 & 	0.0464 & 0.0756 &  & 0.0474 & 0.0832   \\
    & 	     & 	1.00 & 	0.0426 & 0.1032 &  & 0.0462 & 0.1036   \\ \hline
500 & 	0.25 & 	0.25 & 	0.0522 & 0.0630 &  & 0.0456 & 0.0604   \\ 
    & 	     & 	0.50 & 	0.0468 & 0.0926 &  & 0.0510 & 0.1040   \\
    & 	     & 	0.75 & 	0.0496 & 0.1468 &  & 0.0480 & 0.1392   \\
    & 	     & 	1.00 & 	0.0572 & 0.2012 &  & 0.0502 & 0.1910   \\ \hline
    & 	0.50 & 	0.25 & 	0.0596 & 0.0726 &  & 0.0494 & 0.0646   \\ 
    & 	     & 	0.50 & 	0.0494 & 0.1128 &  & 0.0478 & 0.1116   \\
    & 	     & 	0.75 & 	0.0462 & 0.1532 &  & 0.0482 & 0.1590   \\
    & 	     & 	1.00 & 	0.0582 & 0.2152 &  & 0.0506 & 0.2076   \\ \hline
    & 	0.75 & 	0.25 & 	0.0470 & 0.0624 &  & 0.0480 & 0.0650   \\ 
    & 	     & 	0.50 & 	0.0480 & 0.0976 &  & 0.0504 & 0.0940   \\
    & 	     & 	0.75 & 	0.0470 & 0.1488 &  & 0.0506 & 0.1472   \\
    & 	     & 	1.00 & 	0.0474 & 0.1946 &  & 0.0490 & 0.1864   \\ \hline
\end{tabular}

\end{center}

\noindent mNRI test = Modified NRI test with Theorem 3 reference distribution 
\newline NRI test = NRI test with normal reference distribution
\newline $n$ = Sample size within each simulation;  $ \ \ \rho $ = Correlation between covariates $(X,Z)$; $ \ \ \pi_0 = \mbox{Pr}(Y = 1)$;  $ \ \ \mu_X$ = Population mean for $X$ when $Y=1$

\newpage

\begin{center}

{\bf TABLE 2. } Type 1 error for the NRI and the modified NRI
\newline test procedures using a training and an independent test sample

\vskip 0.25in
\renewcommand{\arraystretch}{0.65}

\begin{tabular}{|c|c|c|c|c|c|c|c|}
\multicolumn{3}{c}{ } &
\multicolumn{2}{c}{$\rho = 0$} &
\multicolumn{1}{c}{ } &
\multicolumn{2}{c}{$\rho = 0.5$} \\
\hline

$n$ & $ \pi_0$ &$\mu_X$&  mNRI &  NRI   &  & mNRI   & NRI      \\ 
		&					 &			 &  test &  test  &  & test   & test      \\ \hline
200 & 	0.25 & 	0.25 & 	0.0518 & 0.0492 &  & 0.0490 & 0.0454   \\
    & 	     & 	0.50 & 	0.0500 & 0.0586 &  & 0.0528 & 0.0590   \\
    & 	     & 	0.75 & 	0.0506 & 0.0750 &  & 0.0498 & 0.0718   \\
    &  	     & 	1.00 & 	0.0524 & 0.1128 &  & 0.0468 & 0.1058   \\ \hline
    & 	0.50 & 	0.25 & 	0.0500 & 0.0602 &  & 0.0508 & 0.0556   \\
    & 	     & 	0.50 & 	0.0456 & 0.0722 &  & 0.0534 & 0.0818   \\
    & 	     & 	0.75 & 	0.0496 & 0.0966 &  & 0.0486 & 0.0984   \\
    & 	     & 	1.00 & 	0.0568 & 0.1210 &  & 0.0484 & 0.1288   \\ \hline
    & 	0.75 & 	0.25 & 	0.0516 & 0.0486 &  & 0.0532 & 0.0520   \\ 
    & 	     & 	0.50 & 	0.0496 & 0.0582 &  & 0.0478 & 0.0640   \\
    & 	     & 	0.75 & 	0.0492 & 0.0834 &  & 0.0516 & 0.0834   \\
    & 	     & 	1.00 & 	0.0560 & 0.1076 &  & 0.0484 & 0.1070   \\ \hline
500 & 	0.25 & 	0.25 & 	0.0510 & 0.0620 &  & 0.0528 & 0.0634   \\ 
    & 	     & 	0.50 & 	0.0528 & 0.1060 &  & 0.0494 & 0.0932   \\
    & 	     & 	0.75 & 	0.0542 & 0.1340 &  & 0.0532 & 0.1448   \\
    & 	     & 	1.00 & 	0.0536 & 0.2012 &  & 0.0554 & 0.1862   \\ \hline
    & 	0.50 & 	0.25 & 	0.0594 & 0.0684 &  & 0.0480 & 0.0654   \\ 
    & 	     & 	0.50 & 	0.0518 & 0.1068 &  & 0.0560 & 0.1140   \\
    & 	     & 	0.75 & 	0.0516 & 0.1510 &  & 0.0488 & 0.1528   \\
    & 	     & 	1.00 & 	0.0524 & 0.1924 &  & 0.0464 & 0.1922   \\ \hline
    & 	0.75 & 	0.25 & 	0.0526 & 0.0628 &  & 0.0564 & 0.0610   \\ 
    & 	     & 	0.50 & 	0.0530 & 0.0980 &  & 0.0498 & 0.0998   \\
    & 	     & 	0.75 & 	0.0514 & 0.1476 &  & 0.0500 & 0.1302   \\
    & 	     & 	1.00 & 	0.0504 & 0.1894 &  & 0.0486 & 0.1862   \\ \hline
\end{tabular}

\end{center}

\noindent mNRI test = Modified NRI test with Theorem 4 reference distribution 
\newline NRI test = NRI test with normal reference distribution
\newline $n$ = Sample size within each simulation;  $ \ \ \rho $ = Correlation between covariates $(X,Z)$; $ \ \ \pi_0 = \mbox{Pr}(Y = 1)$;  $ \ \ \mu_X$ = Population mean for $X$ when $Y=1$

\newpage

\begin{center}

{\bf TABLE 3. } NRI and modified NRI for the prostate data
\vskip 0.25in

\renewcommand{\arraystretch}{0.7}  

\begin{tabular}{|l|c|c|c|c|c|}
 \hline

Omitted factor 							&	NRI    	&	P-value 				 	&  mNRI   	& P-value   	\\
														&					&	NRI test					&						& mNRI test	  \\	\hline
Albumin	  									& 0.116  	&	0.236 		 				& 0.018 		& 0.920     	\\
Alkaline phosphatase   			& 0.336   & $<$ 0.001  				& 0.106			& 0.014     	\\
Circulating tumor cells			& 0.627 	&	$<$ 0.001 				& 0.190			& $<$ 0.001    \\  
Gleason score 							& 0.086   & 0.381		 					& 0.034			& 0.849      		\\   
Hemoglobin	  							& 0.351  	&	$<$ 0.001 				& 0.088			& 0.020     		\\
Lactate dehydrogenase	  		& 0.027  	&	0.787 		 				& 0.056			& 0.322     		\\
Prostate specific antigen   & 0.359   & $<$ 0.001 				& 0.080			& 0.138     		\\
Serum testosterone					& 0.195 	&	0.046 		 				& 0.044			& 0.490     	  \\  \hline
\end{tabular}

\end{center}

\mbox{ }

\noindent P-value NRI test = P-value generated from the NRI test procedure with a normal reference distribution 
\newline P-value mNRI test = P-value generated from the mNRI test procedure with the reference distribution specified in Theorem 3.

\newpage 

\centerline{\includegraphics[height=6.6in,width=6.6in]{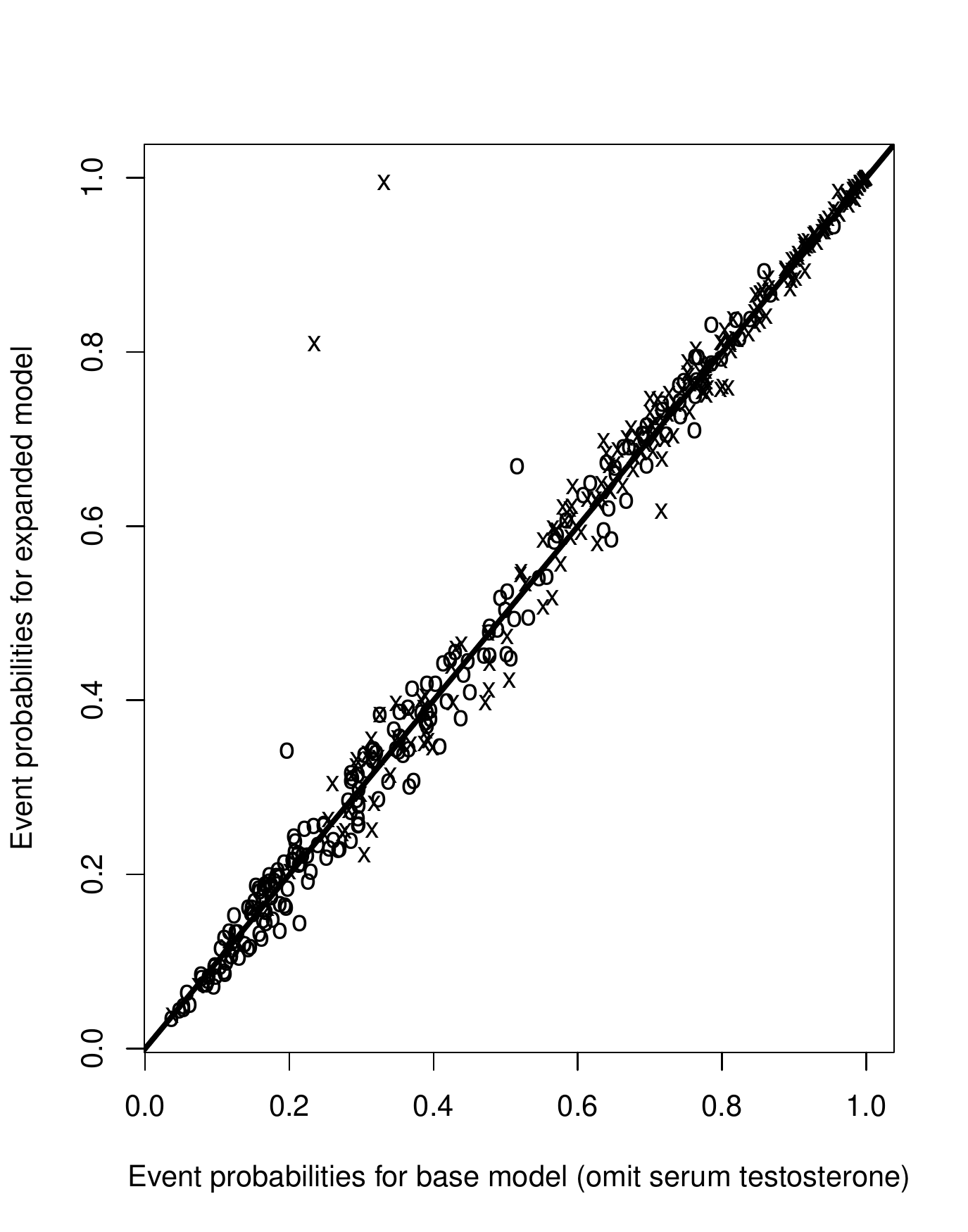}}
\noindent {\bf FIGURE 1 } Event probabilities for each individual estimated from the base model and the expanded model. The expanded model includes all eight biomarkers and the base model omits the biomarker serum testosterone. The symbols 'o' and 'x' represent individuals that survived 24 months from the start of treatment and those who did not.

\newpage

\centerline{\includegraphics[height=6.6in,width=6.6in]{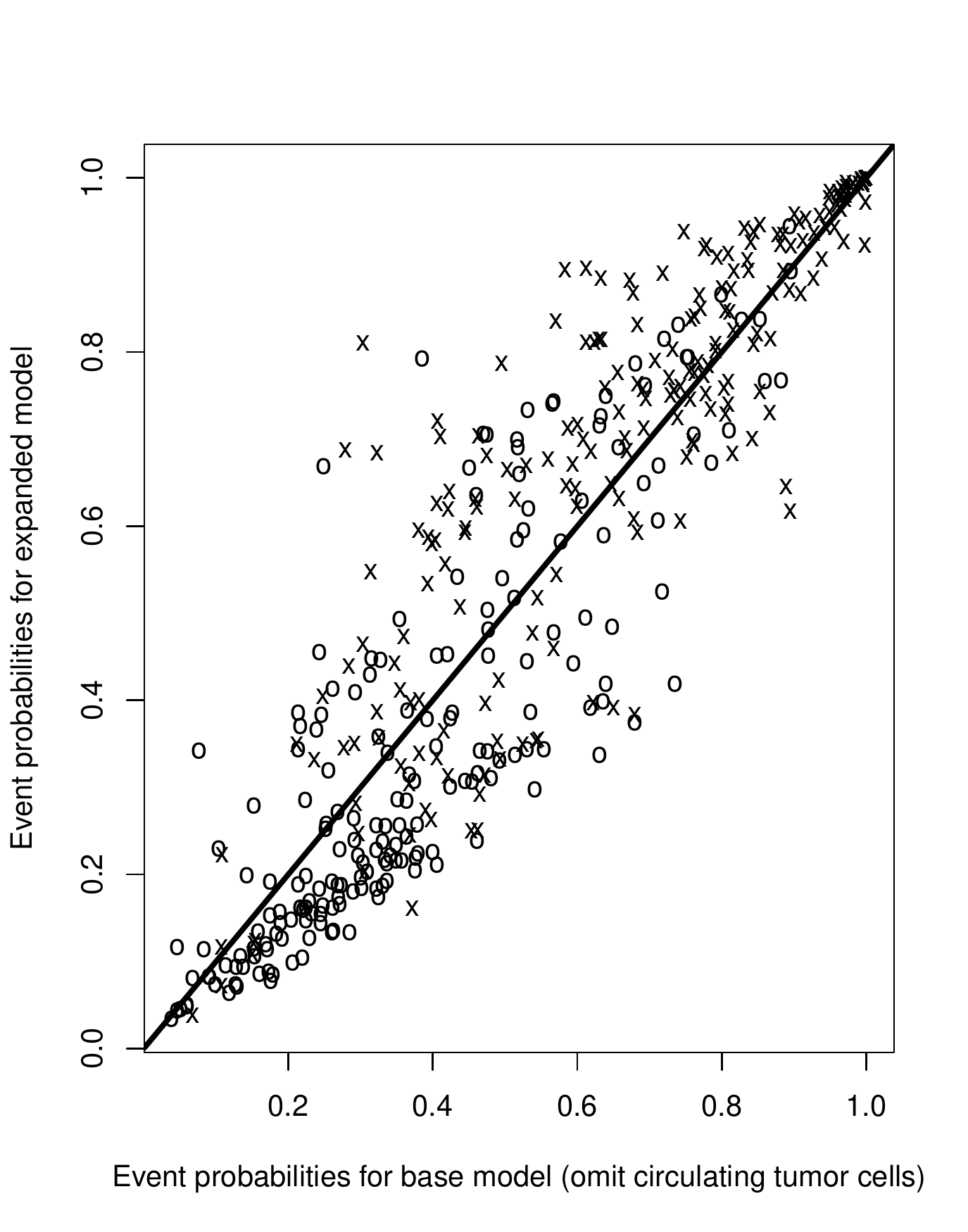}}
\noindent {\bf FIGURE 2 } Event probabilities for each individual estimated from the base model and the expanded model. The expanded model includes all eight biomarkers, and the base model omits the biomarker circulating tumor cells. The symbols 'o' and 'x' represent individuals that survived 24 months from the start of treatment and those who did not.

\end{document}